\documentclass[12pt]{article}
\usepackage{graphicx}

\newcommand\pubnumber{}
\newcommand\pubdate{\today}

\def\hawaii{Department of Physics \& Astronomy\\
University of Hawaii at Manoa, Honolulu, Hawaii}

\def\Title#1{\begin{center} {\Large #1 } \end{center}}
\def\Author#1{\begin{center}{ \sc #1} \end{center}}
\def\Address#1{\begin{center}{ \it #1} \end{center}}

\newcommand\pubblock{\rightline{\begin{tabular}{l} \pubnumber\\
         \pubdate  \end{tabular}}}
\newenvironment{Abstract}{\begin{quotation}  }{\end{quotation}}
\newenvironment{Presented}{\begin{quotation} \begin{center} 
             PRESENTED AT\end{center}\bigskip 
      \begin{center}\begin{large}}{\end{large}\end{center} \end{quotation}}
\def\Acknowledgements{\bigskip  \bigskip \begin{center} \begin{large}
             \bf ACKNOWLEDGEMENTS \end{large}\end{center}}
\def\dcube{D$^3$ }
\def\dcubem{D$^3$-micro }
\def\dcubemilli{D$^3$-milli }

\def\beq{\begin{equation}}
\def\eeq#1{\label{#1}\end{equation}}
\def\eeqn{\end{equation}}

\def\beqa{\begin{eqnarray}}
\def\eeqa#1{\label{#1}\end{eqnarray}}
\def\eeqan{\end{eqnarray}}

\let\bar=\overbar

\def\Dslash{\not{\hbox{\kern-4pt $D$}}}
\def\dslash{\not{\hbox{\kern-2pt $\del$}}}

\def\msb{{\bar{\ssstyle M \kern -1pt S}}}

\begin{document}
\begin{titlepage}
\pubblock

\vfill
\Title{Recent Progress on \dcube \\ The Directional Dark Matter Detector}
\vfill
\Author{Steven Ross on behalf of the \dcube Collaboration}
\Address{\hawaii}
\vfill
\begin{Abstract}
Direction-sensitive WIMP dark matter searches may help overcome the challenges faced by direct dark matter detection experiments. In particular, directional detectors should be able to clearly differentiate a dark matter signal from background sources. We are developing a Directional Dark Matter Detector (D$^3$) based on a gas Time Projection Chamber (TPC) using Gas Electron Multipliers (GEMs) for charge amplification and pixel electronics for readout. This approach allows the three-dimensional reconstruction of nuclear recoils in a room-temperature detector with low energy threshold and low noise. We present an overview of our past and ongoing work developing this technology, including the performance measurements of small prototypes, as well as our planned future work constructing a m$^3$-scale detector to clearly determine whether the signals seen by DAMA, CoGeNT, and CRESST-II are due to low-mass WIMPs or background.
\end{Abstract}
\vfill
\begin{Presented}
CosPA 2013\\
Symposium on Cosmology and Particle Astrophysics\\
Honolulu, Hawaii,  November 12--15, 2013
\end{Presented}
\vfill
\end{titlepage}
\def\thefootnote{\fnsymbol{footnote}}
\setcounter{footnote}{0}

\section{Introduction}

Our group is investigating the feasibility of micro-pattern gas detectors for use in direction-sensitive searches for weakly intreating massive particles (WIMPs). We are specifically evaluating the use of Gas Electron Multipliers (GEMs) and pixel electronics as readout technology for a gas Time Projection Chamber (TPC). The \dcube readout system is one of four candidate technologies for DRIFT-HD, a planned large-scale directional detector to be built as part of the DRIFT (Directional Recoil Identification From Tracks) experiment.

\begin{figure}
\centering
\includegraphics[width=0.7\linewidth]{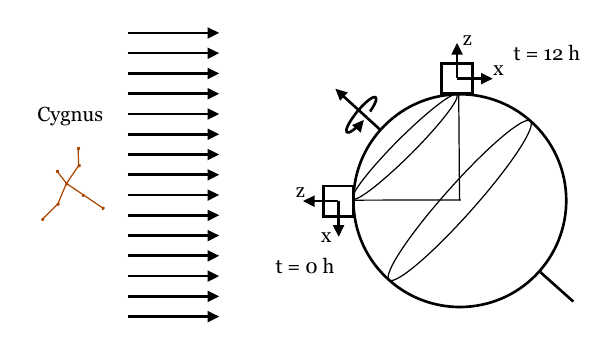}
\caption{The rotation of Earth as it moves through the WIMP wind causes a daily shift in the incoming particle direction as measured in the lab frame \cite{cygnus}.}
\label{dailyosc}
\end{figure}

Directionally-sensitive dark matter detectors are of interest due to their potential to provide unambiguous evidence for the existence of WIMPs. The earth's motion with respect to the galactic rest frame should produce a ``WIMP wind" through the planet originating from the direction of the constellation Cygnus. The rotation of the earth as it moves through this wind would cause a daily 90$^{\circ}$ shift in the incoming particle direction as seen by a detector on the surface. This principle is illustrated in Figure \ref{dailyosc}. There is no known background with a similar signature, making this a very promising detection method. It is also a very large effect, with some models predicting that the signal could be distinguished from background with only $\mathcal{O}$(10) events \cite{copi}.

\section{Current Work at University of Hawaii}

\subsection{Detector Design and Operation}

\begin{figure}
\begin{center}$
\begin{array}{cc}
\includegraphics[width=0.6\linewidth]{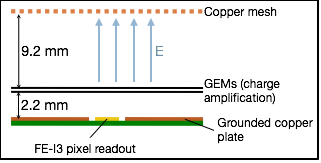} &
\includegraphics[width=0.35\linewidth]{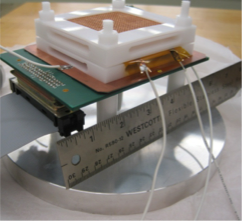}
\end{array}$
\end{center}
\caption{\emph{Left:} Cross-sectional diagram of \dcubem detector (not to scale). \emph{Right:} Image of the \dcubem detector in the lab at University of Hawaii.}
\label{d3cross-sec}
\end{figure}

In 2010, our group built a small prototype detector, \dcubem (shown in Figure \ref{d3cross-sec}). The sensitive volume of the detector (also referred to as the drift gap) is defined by a copper mesh situated 9.2 mm away from the surface of a GEM. Another GEM is located directly below the first, and below that is located the readout plane. This entire structure is then placed inside of a vacuum vessel, which is then filled with gas to be used as the detection medium. 

By holding the mesh and the top GEM at a potential difference, we create an electric field in the drift gap. When ionization is produced in the volume due to particle interactions with the gas, electrons are accelerated by the field and become free electrons. As these electrons drift out of the sensitive volume, they pass through both GEMs, each of which produces avalanche multiplication. The resulting avalanche charge is read out by the ATLAS FE-I3 pixel chip.

The GEMs used are a standard CERN design with a 5 $\times$ 5 cm active area and 140 $\mu$m hole spacing. Using a double GEM layer allows the detector to run at very high gain with low risk of sparking; we have been running this setup in ArCO$_2$ with a gain of $\sim$20,000 at atmospheric pressure. The pixel chip active area is 7 $\times$ 8 mm and consists of 2880 50 $\times$ 400 $\mu$m pixels. Each pixel contains an integrating amplifier, discriminator, shaper, and associated digital controls. This means each pixel can individually measure the time of arrival for drifting charge clusters, and we can freely tune each pixel's threshold. The FE-I3 chip also has very low noise levels -- $\sim$100 electrons per pixel.

The pixel chip is run in self-triggering mode -- that is, no data is communicated from the chip to the DAQ system unless an event is detected. Using the x-y positions of each pixel plus the z-position inferred from time-of-arrival measurement, the FE-I3 should allow for full 3D reconstruction of recoil tracks with head-tail sensitivity. In addition, the combination of high GEM gain and low threshold for the pixels allows for single-electron efficiency, with every electron arriving at the first GEM being read out by the pixel chip.

\subsection{Characterization}

We have performed various studies to characterize the performance of our detector prototype, including measurements of the GEM gain and gain resolution as well as the point and angular resolutions.

\subsubsection{Gain and Gain Resolution}

To study the GEM gain, we used an uncollimated Iron-55 x-ray source, positioned on top of the anode mesh. For this study, we did not use the FE-I3 chip; instead it was replaced by a solid copper pad attached to a pulse height analyzer, allowing us to measure the energy spectrum of our x-ray source. Measurements were taken for two gas mixtures, Ar:CO$_2$ (70:30) and He:CO$_2$ (70:30), both at atmospheric pressure. We have run this setup for weeks at a time at high gain ($\sim10^4$) with no sparking issues.

\begin{figure}
\begin{center}$
\begin{array}{cc}
\includegraphics[width=0.5\linewidth]{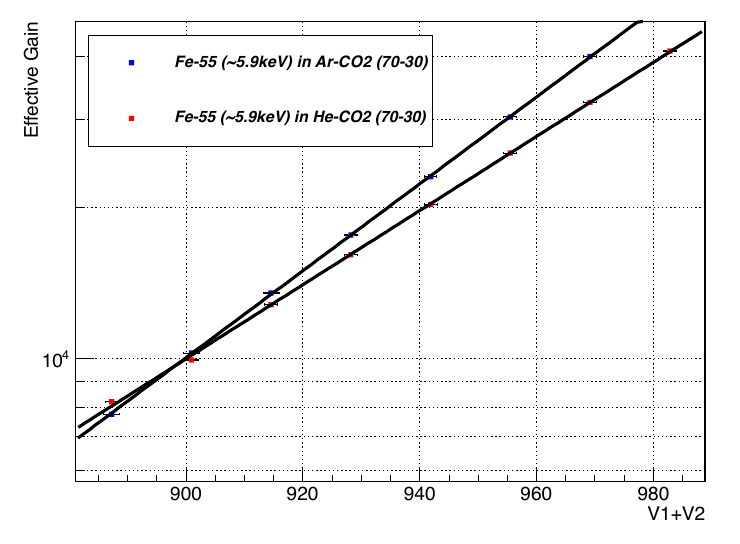} &
\includegraphics[width=0.4\linewidth]{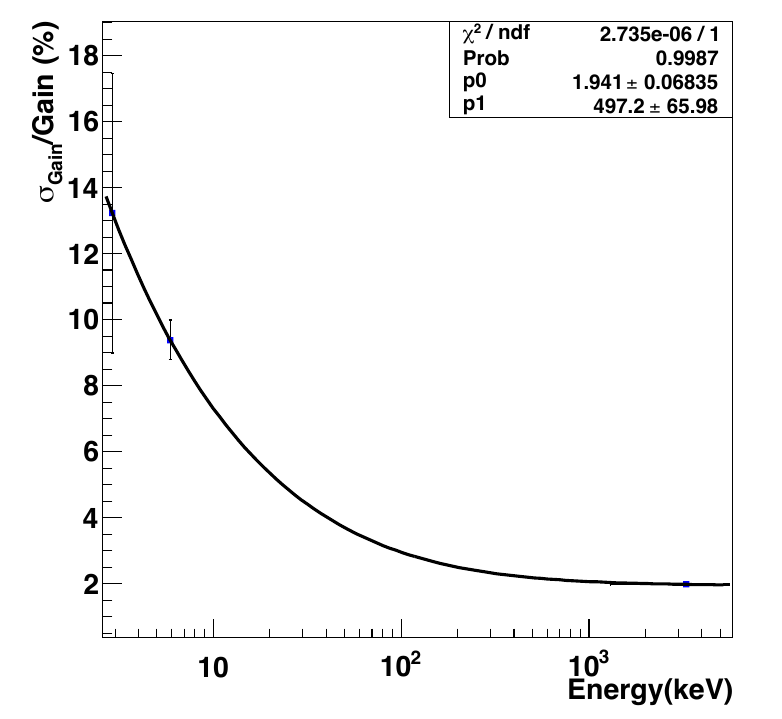}
\end{array}$
\end{center}
\caption{\emph{Left:} Effective gain of the double GEM layer versus the sum of the GEM voltages. \emph{Right:} High-gain resolution plotted against primary ionization energy in ArCO$_2$.}
\label{effectivegain}
\end{figure}

By fitting the energy peaks with a combination of Gaussian and other functions, we are able to determine the gain value (from the Gaussian mean) and gain resolution (from the standard deviation of the Gaussian). Figure \ref{effectivegain} plots the effective gain of our two GEMs versus the sum of the GEM voltages for both gas mixtures. We also studied the asymptotic (high-gain) resolution versus energy in ArCO$_2$, as shown in Figure \ref{effectivegain}. We find that the detector has good gain resolution for MeV-scale signals, and the resolution is adequate even at the keV scale. This indicates that we should achieve good head-tail discrimination with the detector, perhaps even in the keV range.

\subsubsection{Angular Resolution}

\begin{figure}
\centering
\includegraphics[width=0.7\linewidth]{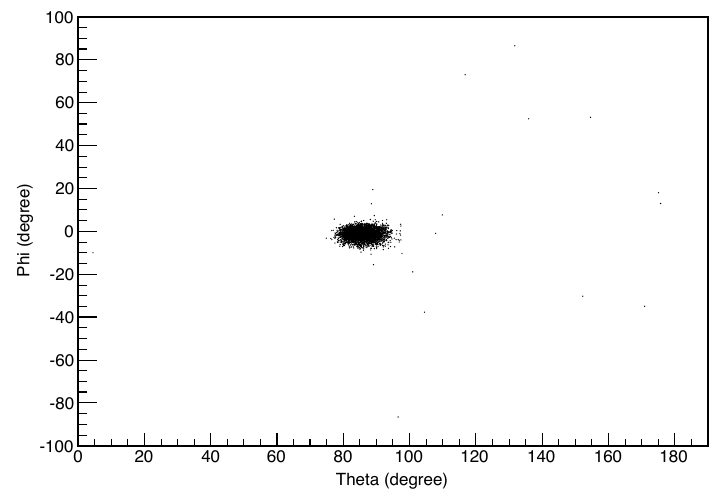}
\caption{Angular distribution of alpha particle tracks produced by Po-210 source.}
\label{angularphitheta}
\end{figure}

\begin{figure}[h]
\begin{center}$
\begin{array}{cc}
\includegraphics{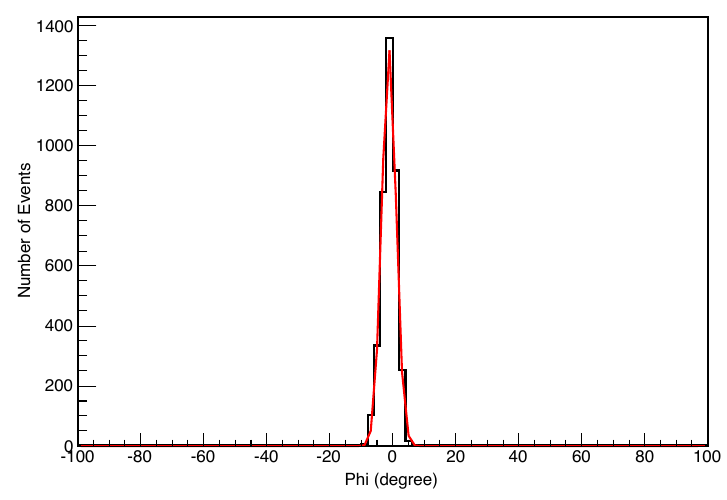} &
\includegraphics{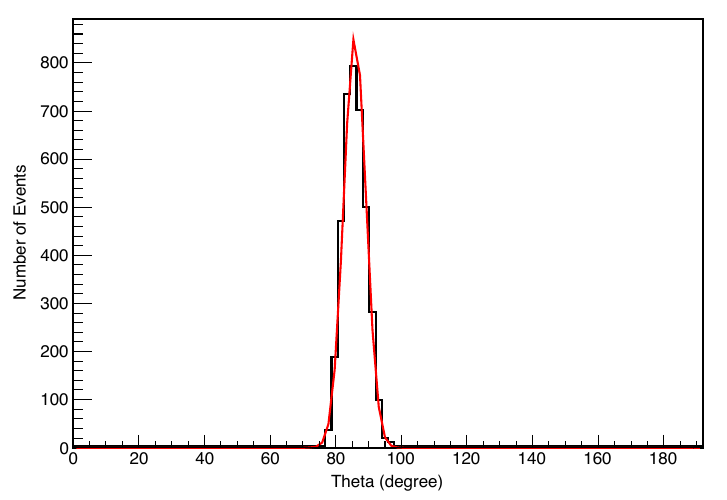}
\end{array}$
\end{center}
\caption{Gaussian fit to the angular distribution of alpha particle tracks in both the polar (\emph{left}) and azimuthal (\emph{right}) directions.}
\label{angulargauss}
\end{figure}

To determine the angular resolution of our detector, we placed a Polonium-210 alpha particle source next to the drift gap, aimed such that the alpha particles traversed the drift region horizontally, above the surface of the pixel chip. We filled the detector with the HeCO$_2$ mixture and looked for tracks of the Helium nuclei. After performing an event selection to obtain a group of well-reconstructed alpha events, we fit each event to a line in 3D space to find its angle of origin. As can be seen in Figure \ref{angularphitheta}, these events clearly point back to a single well-defined source. By fitting a Gaussian curve to the 1D projection of these events in both the polar and azimuthal directions (see Figure \ref{angulargauss}) and determining the standard deviations, we found that the angular resolution was on the order of 1$^{\circ}$ in both directions.

\subsection{Directional Neutron Detection}

\begin{figure}[h]
\begin{center}$
\begin{array}{cc}
\includegraphics[width=0.5\linewidth]{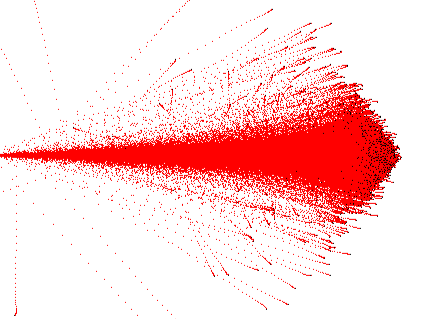} &
\includegraphics[width=0.5\linewidth]{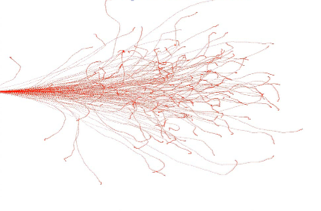}
\end{array}$
\end{center}
\caption{Simulations of nuclear recoils performed with SRIM. \emph{Left:} 1 MeV Hydrogen nuclei recoiling in 1 atm of C$_4$H$_{10}$ gas. \emph{Right:} 100 keV Fluorine nuclei recoiling in 75 Torr of CF$_4$ gas.}
\label{srim}
\end{figure}

With the excellent performance of our prototype detector established with characterization studies, the next goal was to perform directional detection of fast neutrons. Fast neutrons make for a good first demonstration of the dark matter detection principle, because they scatter elastically with the gas, as is expected for WIMPS. Neutron recoils, however, have higher energy and thus are substantially easier to detect. Figure \ref{srim} shows simulations of nuclear recoils caused by neutrons and WIMPs in two different gases.

To demonstrate fast neutron detection, we exposed the detector to a 50 $\mu$C Californium-252 neutron source. The detection medium was atmospheric HeCO$_2$. In initial testing, it was difficult to separate the Helium recoils due to neutrons from background events, caused mainly by radioactivity in our detector materials and protons produced by interactions of the neutrons with the plastic support structure. To counteract these backgrounds, we made the simple modification of increasing the drift gap from 9.2 mm to 5 cm and reduced the amount of material in the acetal support structure, which immediately gave us a dramatically better signal-to-background ratio.

\begin{figure}
\centering
\includegraphics[width=0.7\linewidth]{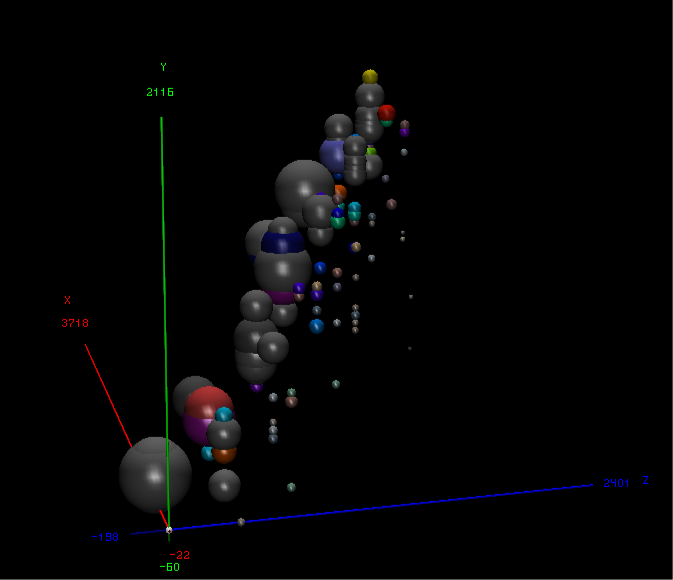}
\caption{Helium nucleus recoil event recorded with modified detector prototype.}
\label{neutron}
\end{figure}

\begin{figure}[h]
\begin{center}$
\begin{array}{cc}
\includegraphics[width=0.5\linewidth]{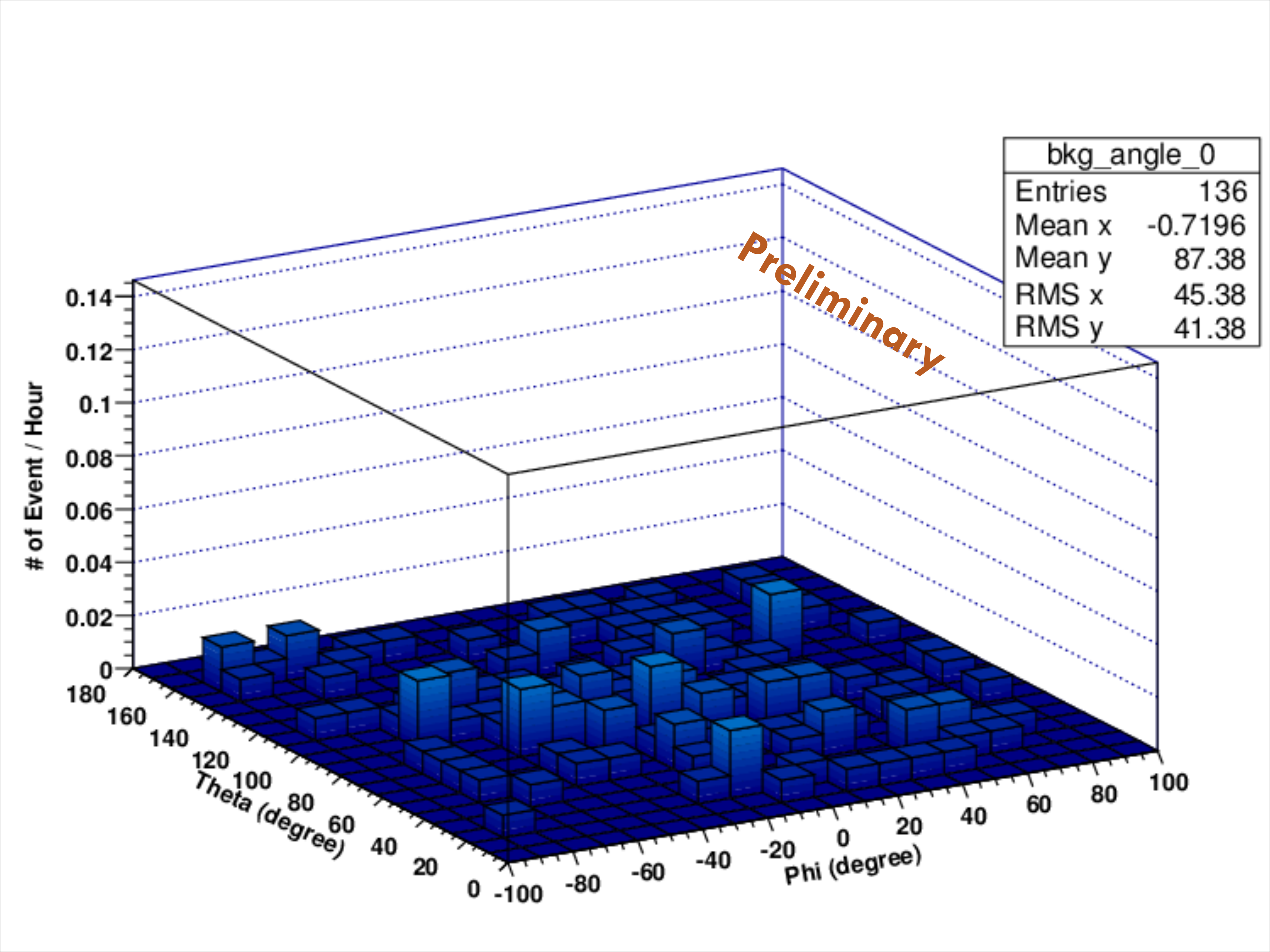} &
\includegraphics[width=0.5\linewidth]{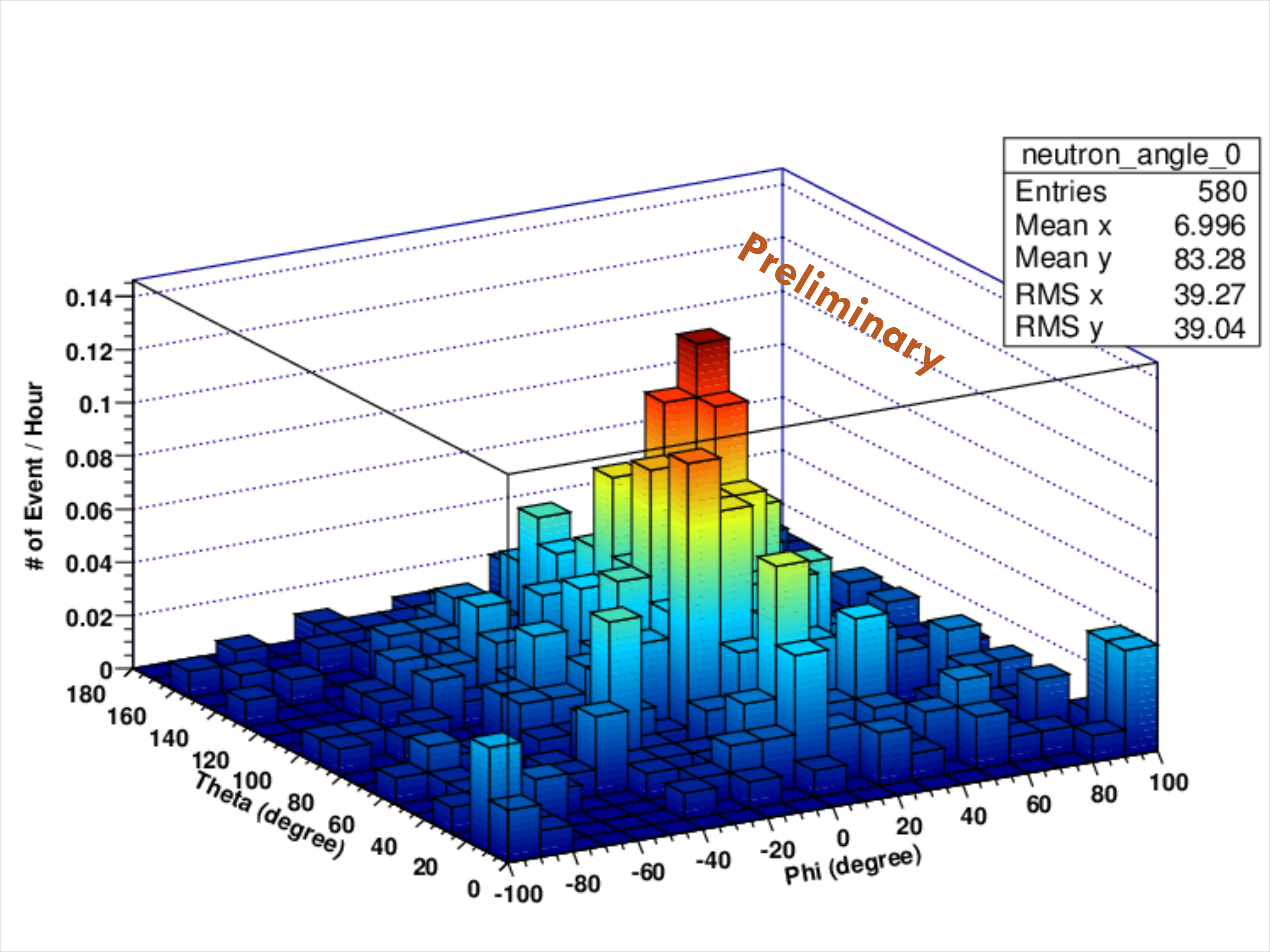}
\end{array}$
\end{center}
\caption{Comparison of two sets of events recorded with modified detector prototype. The set taken with no neutron source present (\emph{left}) has a relatively flat distribution, while the one taken with the the Cf-252 source pointing at the detector (\emph{right}) shows an obvious peak in the direction of the neutron source.}
\label{neutronruns}
\end{figure}

With this modified design, the detector clearly and correctly pointed back to the neutron source (an example neutron event is shown in Figure \ref{neutron}). Figure \ref{neutronruns} shows two data sets, one taken with the neutron source pointing at the detector and the other with the source absent. At this point, no particle identification or event selection cuts have been applied to the data, but we are already able to clearly identify the source direction. We are currently working to implement particle identification cuts as well as head-tail discrimination to improve the directionality of the detector.

\section{Application of Technology to WIMP Detection}

We have now demonstrated precise, 3D directional nuclear recoil detection in the 100-500 keV range with a small TPC. However, some questions remain as we look to apply this technology to WIMP detection. One uncertainly is whether the detector will remain effective for low energy recoils (in the range of 10 keV). Sensitivity to these events is important for exploring the possibility of low mass WIMPs. With our upcoming detector prototypes, we will investigate factors important for detection at low energies, including the operation of the detector at low gas pressures and the physical characteristics of low energy nuclear recoils (e.g. quenching factor and straggling).

\begin{figure}
\centering
\includegraphics[width=0.7\linewidth]{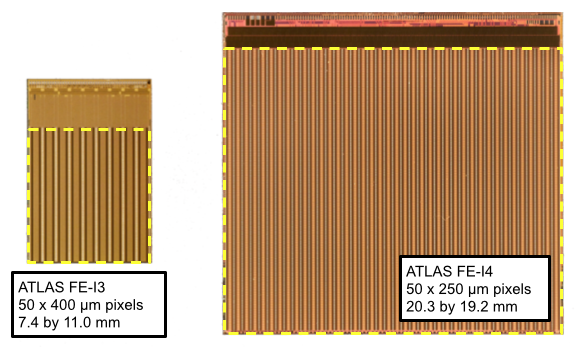}
\caption{Side-by-side comparison of the ATLAS FE-I3 and FE-I4 pixel chips \emph{Left:} The FE-I3 uses an array of 50 $\times$ 400 $\mu$m pixels, with a 7.4 $\times$ 11.0 mm active area. \emph{Right:} the FE-I4 uses 50 $\times$ 250 $\mu$m pixels and has an active area of 20.3 $\times$ 19.2 mm, making it both larger and higher in resolution than the FE-I3.}
\label{chipcompare}
\end{figure}

To reach the target mass necessary for competitive WIMP detection, it is also important that we be able to build large-scale TPCs. We are currently exploring a few options to help limit the cost per unit volume of our detectors. The first is an upgrade to the next-generation ATLAS pixel chip, the FE-I4. The active area of the new chip is approximately five times the size of our current FE-I3 chip, which would allow us to instrument a larger readout volume at a lower total electronics cost. A comparison of the two pixel chips is shown in Figure \ref{chipcompare}. Our collaborators at Lawrence Berkeley National Laboratory have installed the FE-I4 in their TPC and run initial tests, while the Hawaii group is in the process of building a new detector prototype using the next-generation chip.

We are also developing the charge-focusing readout of TPCs, a novel method that would allow a small area of electronics to be sensitive to a much larger detector volume \cite{ross}. By focusing the drift charge with electrostatic fields, we would increase the detector volume read out by the pixel chip while maintaining its small feature size, which is important for limiting noise levels. Simulations suggest that we should be able to achieve homogeneous focusing across the readout area while limiting the diffusion, so that we would be able to retain the directionality of recoil tracks. The first experimental tests of this idea yielded promising results, but a more detailed analysis needs to be done to fully establish its feasibility.

\section{Conclusions and Outlook}

The results of the \dcubem prototype show that a TPC using GEMs and pixel readout can achieve precise, 3D directional detection of nuclear recoils. The detector can be run stably for extended periods, and the GEMs allow for high gain with good resolution, which should enable accurate head-tail identification for recoils. Overall, the results are promising for the application of this detector technology to the search for WIMPs.

\begin{figure}
\centering
\includegraphics[width=0.7\linewidth]{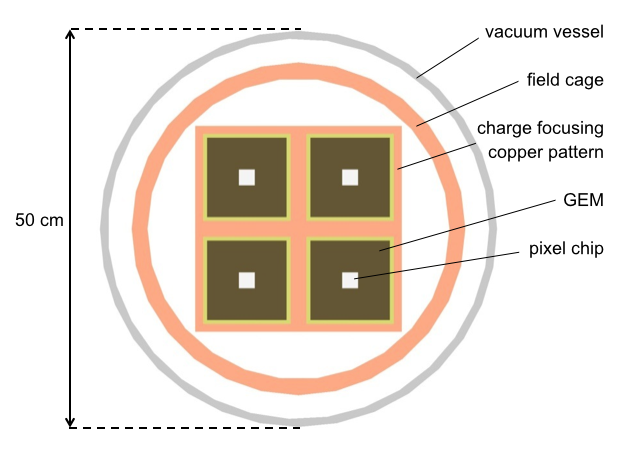}
\caption{Top-down view of \dcubemilli design, which implements four unit cells inside a common field cage.}
\label{dcubemilli}
\end{figure}

In the immediate future, we will begin work on a larger prototype detector, \dcubemilli (depicted in Figure \ref{dcubemilli}). This detector will be much larger than the current \dcubem, with approximately 10 liters of active volume. It will use the new FE-I4 pixel chip along with 10 $\times$ 10 cm GEMs from CERN, as well as an upgraded DAQ system.

We have also recently joined the DRIFT Experiment, the world's leading directional dark matter search effort. DRIFT is the only experiment running a directionally-sensitive detector on the m$^3$ scale, and they are currently working on plans for DRIFT-III, a 24 m$^3$ detector with 4 kg of target mass. We believe that combining our readout technology with DRIFT's experience and expertise will give us a great advantage in building a competitive directionally-sensitive WIMP detector.

\Acknowledgements
We acknowledge fruitful discussions with John Kadyk and Maurice
Garcia-Sciveres of Lawrence Berkeley National Laboratory, and thank
them for providing the ATLAS FE-I3 pixel chip and associated readout
electronics used in this work. We acknowledge support from the U.S.
Department of Homeland Security under Award Number
2011-DN-077-ARI050-03 and the U.S. Department of Energy under Award
Number DE-SC0007852. The views and conclusions contained in this
document are those of the authors and should not be interpreted as
necessarily representing the official policies, either expressed or
implied, of the United States Government or any agency thereof.


\begin{thebibliography}{99}


\bibitem{cygnus}
S. Ahlen \emph{et. al.}, International Journal of Modern Physics A {\bf 25}, 1 (2010)

\bibitem{copi}
C. J. Copi, J. Heo, and L. M. Krauss, Physics Letters B {\bf 461}, 43 (1999)

\bibitem{ross}
S. J. Ross \emph{et. al.},  2012 IEEE Nuclear Science Symposium and Medical Imaging Conference Record, 1760 (2012)


\end{thebibliography}
\end{document}